\documentclass[12pt, prd, showpacs]{revtex4}
\usepackage{amssymb}
\usepackage{amsmath}

\input{tcilatex}

\begin{document}

\title{The Ba\~{n}ados-Silk-West effect with immovable particles near static
black holes and its rotational counterpart}
\author{O. B. Zaslavskii}
\affiliation{Department of Physics and Technology, Kharkov V.N. Karazin National
University, 4 Svoboda Square, Kharkov 61022, Ukraine}
\email{zaslav@ukr.net}

\begin{abstract}
The BSW effect implies that the energy $E_{c.m.}$ in the center of mass
frame of two particles colliding near a black hole can become unbounded.
Usually, it is assumed that particles move along geodesics or
electrogeodesics. Instead, we consider another version of this effect. One
particle is situated at rest near a static, generally speaking, distorted
black hole. If another particle (say, coming from infinity) collides with
it, the energy of collision $E_{c.m.}$ in the center of mass frame grows
unbounded (the BSW effect). The force required to keep such a particle near
a black hole diverges for nonextremal horizons but remains finite nonzero
for extremal one and vanishes in the horizon limit for ultraextremal black
holes. Generalization to the rotating case implies that a particle corotates
with a black hole but does not have a radial velocity. In doing so, the
energy $E\rightarrow 0$, provided the angular momentum $L=0$. This condition
replaces that of fine-tuning parameters in the standard version of the BSW
effect.
\end{abstract}

\keywords{}
\pacs{04.70.Bw, 97.60.Lf }
\maketitle

\section{Introduction}

High energy particle collisions near black holes attract much attention
after findings of Ba\~{n}ados, Silk and West who showed that collision of
free falling particles in the background of the extremal rotating black hole
can produce an indefinitely large energy $E_{c.m.}$ in the center of mass
frame \cite{ban} (the BSW effect, after the names of its authors). This
requires one of particles to be fine tuned in such a way that the Killing
energy and angular momentum obey the relation 
\begin{equation}
X\equiv E-\omega _{H}L=0,  \label{cr}
\end{equation}%
$\omega _{H}$ being the angular velocity of a black hole (this is so-called
critical particle). Thus a fine-tuned (so-called critical) particle collides
with a usual (not fine-tuned) one to produce the unbounded growth of $%
E_{c.m.}$,

This is what can be called the basic scenario of the BSW effect. Within this
type of the scenario, the BSW effect is impossible for static black holes
since $\omega _{H}=0$ for them, and the equality (\ref{cr}) fails.
Meanwhile, there exists another, special type auxiliary to the basic one.
Within this special type, the BSW could be possible, in principle, even near
static black holes, say, the Schwarzschild one. Let $L=0$. Then, (\ref{cr})
requires that $E=0$ or $E$ should be arbitrarily close to $0$. This special
types can be decomposed to the following subcases. (i) Particle with $%
E\approx 0$ moves freely and collides with a usual particle in the vicinity
of the Schwarzschild black hole \cite{eva}, (ii) particle with $E\approx 0$
is kept fixed near the Schwarzschild horizon and collides with a usual
particle falling from infinity \cite{gp11} (see discussion on p. 3864 before
Eq. (35) there). However, each of both subcases encounters serious
difficulties that prevent if from physical realization. As far as subcase
(i) is concerned, it was shown later that such a particle cannot come from
infinity and, moreover, it cannot be created near a black hole as a result
of previous collisions between particles falling from infinity \cite{spec}.
In subcase (ii) there is a difficulty of another kind: a force required to
keep a particle near the horizon grows indefinitely.

The aim of the present work is to show that there is one more type of a
special scenario free of the aforementioned difficulties. In this sense, we
fill the gap in the classification of the BSW scenarios left before, and
bring this issue to completion. We also demonstrate the analogue of the
special type scenario for rotating black holes. Thus we expand the class of
objects and scenarios for which high energy collisions are possible.

There is one more aspect of the issue under discussion. Some other cursory
notes were made in literature on the BSW effect concerning the states with $%
E=0$. The criticality condition with $E=0$ was considered in Sec. II of \cite%
{ots} for 2 + 1 black holes with the cosmological constant \cite{btz}. It
appeared as a limit of the angular momentum $L\rightarrow 0$ in a more
general condition (2.15) there. The corresponding space-time is not
asymptotically flat but near the horizon the same features manifest
themselves, so $E_{c.m.}$ grows unbounded when $E\rightarrow 0$ for one of
two particles. Earlier, it was also pointed out (see, e.g. , \cite{kd}, page
4) that the values $E=0$ and $L=0$ are possible but these special cases were
mentioned in the context where freely moving particles were considered.
Meanwhile, we show that these cases are relevant for \textit{nongeodesic }%
particles which are kept in equilibrium by some force.

\section{Collisions: basic formulas}

Let two particles 1 and 2 collide in some point. Then, the energy $E_{c.m.}$
in the center of mass frame is defined according to%
\begin{equation}
E_{c.m.}^{2}=-(m_{1}u_{1}^{\mu }+m_{2}u_{2}^{\mu })(m_{1}u_{1\mu
}+m_{2}u_{2\mu })=m_{1}^{2}+m_{2}^{2}+2m_{1}m_{2}\gamma \text{.}
\end{equation}%
Here, $m_{i}$ are masses, $u_{i}^{\mu }$ are four-velocities, $i=1,2,$%
\begin{equation}
\gamma =-u_{1\mu }u_{2}^{\mu }  \label{ga}
\end{equation}%
is the Lorentz factor of relative motion.

\section{Behavior of acceleration. Static case}

For what follows we will need to know the behavior of the acceleration near
the horizon. Our main goal is to show that a finite acceleration can be
reconciled with the BSW effect. To this end, we enumerate all possible
variants. In doing so, we extend formulas well known for spherically
symmetric cases (see, e.g. \cite{fn} for the Schwarzschild case) to more
general metrics and other types of a horizon.

Let us consider the gravitational field described by the static black hole
metric \thinspace $g_{\mu \nu }$. We assume no spatial symmetry. It can be
represented (at least in some finite region) in the form%
\begin{equation}
ds^{2}=-N^{2}dt^{2}+dn^{2}+\gamma _{ab}dx^{a}dx^{b}\text{,}
\end{equation}%
$a,b=1,2$. The lapse function $N=0$ on the horizon. Here, the quantity $n$
measures the proper distance to the horizon. Thus this is, in general, a
distorted black hole.

Let a particle be at rest, having $n=n_{0}=const$, $x^{a}=const$. Then, it
follows from the normalization condition $u_{\mu }u^{\mu }=-1$ for the
four-velocity that $u^{0}=1/N$, $u_{0}=-N$. It is easy to calculate the
components of acceleration $a^{\mu }$: $a^{0}=0$,%
\begin{equation}
a^{n}=\Gamma _{00}^{n}\left( u^{0}\right) ^{2}=a^{n}=\frac{1}{N}\frac{%
\partial N}{\partial n}\text{,}  \label{an}
\end{equation}%
\begin{equation}
a^{a}=\frac{1}{N^{2}}\Gamma _{00}^{a}=\frac{1}{N}\frac{\partial N}{\partial
x^{b}}\gamma ^{ab}\text{,}
\end{equation}%
where $\gamma ^{ab}$ is the tensor inverse to $\gamma _{ab}.$

Now, we consider three cases separately depending on the near-horizon
behavior of the metric. In particular, the crucial role is played here by
the surface gravity $\kappa $ defined according to%
\begin{equation}
\kappa =\lim_{N\rightarrow 0}\sqrt{(\nabla N)^{2}}\text{,}  \label{ka}
\end{equation}%
where $(\nabla N)^{2}=g^{\mu \nu }N_{,\mu }N_{,\nu }$.

\subsection{Nonextremal black hole}

For the nonextremal black hole, the conditions of the regularity near the
horizon require \cite{vis}%
\begin{equation}
N=\kappa n+b(x^{a})n^{3}+o(n^{3})\text{.}
\end{equation}%
We see that in the horizon limit when $n\rightarrow 0$, $N\rightarrow 0$%
\begin{equation}
a^{n}\approx \frac{1}{n}\text{, }a^{a}=O(n^{2})
\end{equation}%
\begin{equation}
a^{2}=a_{\mu }a^{\mu }\approx \frac{1}{n^{2}}\text{,}
\end{equation}%
so $a^{2}$ diverges. This is direct generalization of the situation in the
spherically symmetric static gravitational field (see eq. 2.2.6 in \cite{fn}%
).

\subsection{Extremal black hole}

By definition, this means that near the horizon%
\begin{equation}
N\approx N_{0}\exp (-cn)\text{,}
\end{equation}%
where $N_{0}$ and $c$ are some constants \cite{tz},%
\begin{equation}
\frac{\partial \gamma _{ab}}{\partial n}=O\exp (-cn).
\end{equation}

The horizon limit $N\rightarrow 0$ is achieved when $n\rightarrow \infty $.
As a result, $a^{2}\rightarrow c^{2}$ and remains bounded and nonzero.

\subsection{Ultraextremal black hole}

By definition, the horizon is called ultraextremal if in the horizon limit $%
n\rightarrow \infty $, 
\begin{equation}
N\approx \frac{N_{0}}{n^{s}}  \label{s}
\end{equation}%
with $s>0$, $N_{0}>0$ being some constant. This definition becomes more
clear in the particular case of a spherically symmetric metric%
\begin{equation}
ds^{2}=-fdt^{2}+\frac{dr^{2}}{f}+r^{2}(d\theta ^{2}+\sin ^{2}\theta d\phi
^{2})\text{,}
\end{equation}%
where $f=f(r).$ If 
\begin{equation}
f\approx f_{0}(r-r_{+})^{k}\text{,}  \label{k}
\end{equation}%
where $r_{+}$ is the horizon radius, $f_{0}$ is constant, $k=3,4,5..$.
Comparing (\ref{s}) and (\ref{k}), we see that in this case $s=\frac{k}{k-2}$%
. The regularity coniditons of metrics of this type were studied in Refs. 
\cite{vis}, \cite{tz}.

Now,%
\begin{equation}
a\approx \frac{s}{n}\rightarrow 0\text{.}
\end{equation}

In all three cases the energy per unit mass $\varepsilon =\frac{E}{m}$ is
equal to 
\begin{equation}
\varepsilon =-u_{0}=N\text{.}
\end{equation}%
We see that in the horizon limit $\varepsilon \rightarrow 0$ thus realizing
the special case of the critical condition (\ref{cr}) for static black holes
($\omega _{H}=0$).

\section{Behavior of acceleration. Stationary case}

Let us consider a metric of the stationary axially symmetric black hole. It
can be written in the form%
\begin{equation}
ds^{2}=-N^{2}dt^{2}+g_{\phi }(d\phi -\omega dt)^{2}+\frac{dr^{2}}{A}%
+g_{\theta }d\theta ^{2}\text{,}
\end{equation}%
where all metric coefficients do not depend on $t$ and $\phi $.

Let a particle move over circle with $u^{r}=0=u^{\theta }$%
\begin{equation}
u^{\mu }=u^{t}(1,\Omega ,0,0)\text{,}
\end{equation}%
where%
\begin{equation}
\Omega =\frac{d\phi }{dt}.
\end{equation}

From the condition that a trajectory cannot be space-like, $ds^{2}\leq 0$ we
obtain%
\begin{equation}
\omega _{-}\leq \Omega \leq \omega _{+}
\end{equation}%
where%
\begin{equation}
\omega _{\pm }=\omega \pm \frac{N}{\sqrt{g_{\phi }}},
\end{equation}%
\begin{equation}
\Omega =\omega +\frac{N}{\sqrt{g_{\phi }}}\alpha ,
\end{equation}%
where $\left\vert \alpha \right\vert \leq 1$. It follows from $g_{\mu \nu
}u^{\mu }u^{\nu }=-1$ that%
\begin{equation}
u^{t}=\frac{1}{\sqrt{N^{2}-g_{\phi }(\omega -\Omega )^{2}}}=\frac{1}{N\sqrt{%
1-\alpha ^{2}}}\text{,}  \label{ut}
\end{equation}%
\begin{equation}
u^{\phi }=\Omega u^{t}=\frac{\Omega }{N\sqrt{1-\alpha ^{2}}}=\frac{\omega +%
\frac{N}{\sqrt{g_{\phi }}}\alpha }{N\sqrt{1-\alpha ^{2}}}\text{.}
\end{equation}%
We also have for the angular momentum $L$ and energy $E$%
\begin{equation}
\mathcal{L}\equiv \frac{L}{m}=u_{\phi }=g_{\phi }\frac{(\Omega -\omega )}{N%
\sqrt{1-\alpha ^{2}}}=\frac{\sqrt{g_{\phi }}\alpha }{\sqrt{1-\alpha ^{2}}},
\end{equation}%
\begin{equation}
\varepsilon =\frac{E}{m}=-u_{t}=\frac{1}{\sqrt{1-\alpha ^{2}}}(\omega \sqrt{%
g_{\phi }}\alpha +N).
\end{equation}%
\begin{equation}
\frac{X}{m}\equiv \mathcal{X}=\varepsilon -\omega \mathcal{L}=\frac{N}{\sqrt{%
1-\alpha ^{2}}}\text{.}  \label{x}
\end{equation}

For simplicity, we choose $\alpha =0$. This corresponds to a so-called
zero-angular momentum observer (ZAMO), $L=0$ \cite{72}.

In the horizon limit, rotation is inevitable there due to a strong
frame-dragging effect. Now, instead of being at rest, a particle should
corotate with a black hole. In doing so, $X\rightarrow 0$, so this is a
so-called critical particle according to standard nomenclature \cite{prd}.

\subsection{Particles near rotating black holes}

Let us consider collision between particle 1 rotating around a black hole as
described above and particle 2 moving freely. For simplicity, we choose
particle 2 to move within the equatorial plane.

Then, for particle 2 we have from equations of motion 
\begin{equation}
u_{2}^{\mu }=(\frac{\mathcal{X}_{2}}{N^{2}}\text{, }\frac{\mathcal{L}}{%
g_{\phi }}+\frac{\omega \mathcal{X}_{2}}{N^{2}}\text{, }-\frac{\sqrt{A}}{N}%
\sqrt{\mathcal{X}_{2}^{2}-N^{2}(1+\frac{\mathcal{L}_{2}^{2}}{g_{\phi }})}%
\text{, }0).
\end{equation}

Then, we obtain from (\ref{ga})%
\begin{equation}
\gamma =-u_{2}^{\mu }\left( u_{\mu }\right) _{1}=\frac{\mathcal{X}_{2}}{N}.
\end{equation}%
In the limit $N\rightarrow 0$ we see that $\gamma \sim \frac{1}{N}%
\rightarrow \infty $. Thus the BSW does occur.

Now, we calculate the components of acceleration. For particle 1 with $%
L=0\,\ $\ one can find easily $a^{t}=0$.%
\begin{equation}
a^{r}=\frac{A}{N}\partial _{r}N\text{,}
\end{equation}%
where (\ref{ut}) with $\alpha =0$ was used. Then,%
\begin{equation}
a^{2}=\frac{A}{N^{2}}\left( \partial _{r}N\right) ^{2}.
\end{equation}%
Near the horizon, we assume $A\sim N^{2}$. This is an additional assumption
but it holds for the most metrics of physical interest. In particular, it is
true for the Kerr and Kerr-Newman metrics. (As one can make rescaling of the
radial coordinate, this can be violated far from the horizon but we are
interested in the vicinity of the horizon only.) In the static limit it
holds for the Schwarzschild and Reissner-Nordstr\"{o}m ones everywhere. For
the nonextremal black holes $N\sim \sqrt{r-r_{+}}$, for extremal black holes 
$N\sim r-r_{+}$, for ultraextremal ones $N\sim (r-r_{+})^{p}$, with $p>1$.
Using a general formula (\ref{ka}), it is easy to confirm that $\kappa \neq
0 $ in the first case and $\kappa =0$ in the second and third ones.

Thus in the horizon limit $a^{2}$ diverges for the nonextremal case, is
finite nonzero in the extremal case and vanishes in the ultraextremal one.

\section{Limiting transition and fake orbits on horizon}

Thus we saw that for the extremal and ultraextremal horizons one can keep a
particle as close to the horizon as one likes. In the first case the force
remains a finite nonzero, in the second case it tends to zero. It is natural
to ask: is it possible to bring the limiting procedure to completion and
keep a particle on the horizon itself? The answer is negative. In doing so,
the fact that trajectories of particles on the horizon are fake, reveals
itself. This is because a problem is connected not only with a dynamics but
also with kinematics. As a horizon is a light-like surface, the time-like
trajectories of massive particles are impossible on it. Although formally
one can put $r=r_{+}$, the corresponding trajectory is forbidden. The
situation is already described in details in Sec. III C of Ref. \cite%
{innermost} (see also brief discussion after eq. 4.2 in \cite{kd}) and Sec.
VI of Ref. \cite{circ-15}.

\section{Discussion and Conclusions}

Thus we showed that, indeed, there is a scenario in which (i) a particle
remains in rest near the horizon (in the static case) or corotates with a
black hole (stationary case), (ii) as a consequence, its energy is zero in
this limit, (iii) the acceleration experienced by a particle is finite
nonzero for the extremal black hole or zero for the ultraextremal one, (iv)
its collision with infalling particles leads to an indefinite growth of the
energy in the center of mass frame.

It was explained earlier that, kinematically, the BSW effect arises due to
collision between a slow target and rapid particle that hits it \ \cite{k}.
Now, this circumstance reaches its ultimate form for static black holes
since one of particle does not move before collision at all.

It is worth noting that a role of a particle can also be played by a
macroscopic body, being slowly lowered by a rope towards a black hole.
Models of such a kind are used time to time in black hole physics. In
particular, they were found to be instructive in the discussion of
Gedankenexperiments with thermal radiation (see pp. 479 - 489 in \cite{fn}
and references therein). Now, as we see, this model appeared in a quite
different context.

There is one more aspect. Deviation from geodesic motion was pointed out as
one of potential restrictions acting against the BSW effect \cite{berti}.
Meanwhile, the present work clearly shows that, by itself, the presence of a
force can be compatible with the BSW effect thus realizing, as a simple
example, a general scheme suggested in \cite{fe}.

The scenarios studied in the present work can be considered not only on its
own right but also as a qualitative approximation for more complicated
processes in the near-horizon plasma, when particles move, say, under the
action of the electromagnetic force, being also bombarded by colliding
particles.

\section{Acknowledgement}

I thank for hospitality Vojtech Pravda and Institute of Mathematics in
Prague where this work was conceived during my short visit. I also
acknowledge Funda\c{c}\~{a}o para a Ci\^{e}ncia e Tecnologia - FCT,
Portugal, for financial support through Project No.~UIDB/00099/2020.

\end{document}